\documentclass[12pt]{iopart}   
\usepackage{graphicx}
\usepackage{epsfig}
\usepackage{amssymb}
\usepackage{array}
 
\newcommand{\be}{\begin{equation}}
\newcommand{\ee}{\end{equation}}
\newcommand{\beqn}{\begin{eqnarray}}
\newcommand{\eeqn}{\end{eqnarray}}
\begin{document}

\title[Entanglement Entropy of Disordered Quantum Wire Junctions]{Entanglement Entropy of Disordered Quantum Wire Junctions}
\author{R\'obert Juh\'asz}
\address{Wigner Research Centre for Physics, Institute for Solid State
Physics and Optics, H-1525 Budapest, P.O.Box 49, Hungary}
\ead{juhasz.robert@wigner.mta.hu} 
\author{Johannes M. Oberreuter}
\address{Data Reply GmbH, Luise-Ullrich-Straße 14, 80636 München, Germany}
\address{Department of Physics and Institute for Advanced Study, Technical University of Munich, 85748 Garching, Germany}
\author{Zolt\'an Zimbor\'as}
\address{Wigner Research Centre for Physics, Institute for Particle and Nuclear Physics, H-1525 Budapest, P.O.Box 49, Hungary}
\ead{zimboras.zoltan@wigner.mta.hu}

\date{\today}

\begin{abstract}
We consider different disordered lattice models  composed of $M$ linear chains glued together in a star-like manner, and study the scaling of the entanglement between one arm and the rest of the system using a numerical strong-disorder renormalization group method. For all studied models, the random transverse-field Ising model (RTIM), the random XX spin model, and the free-fermion model with random nearest-neighbor hopping terms, the average entanglement entropy is found to increase with the length $L$ of the arms according to the form $S(L)=\frac{c_{\rm eff}}{6}\ln L+const$. For the RTIM and the XX model, the effective central charge $c_{\rm eff}$ is universal with respect to the details of junction, and only depends on the number $M$ of arms. Interestingly, for the RTIM $c_{\rm eff}$ decreases with $M$, whereas for the XX model it increases. For the free-fermion model, $c_{\rm eff}$ depends also on the details of the junction, which is related to the sublattice symmetry of the model. In this case, both increasing and decreasing tendency with $M$ can be realized with appropriate junction geometries.      
\end{abstract}

\maketitle

\section{Introduction}
\label{sec:intro}

The entanglement properties of extended quantum systems have attracted much interest in the recent decade \cite{amico,entanglement_review, area, laflorencie}.  One reason for this is that various entanglement measures turned out be sensitive to whether the underlying model is critical or not, moreover, some of these showed universal scaling in critical points. For a subsystem $A$ of a closed system in a pure state, the natural entanglement measure is the entanglement entropy $S_A$, which is the von Neumann entropy of the reduced density matrix $\rho_A$ corresponding to the subsystem: 
\be  
S_A=-{\rm tr}(\rho_A\ln\rho_A). 
\label{ee}
\ee
In non-critical models, $S_A$ is proportional to the boundary area of subsystem $A$, while in critical models corrections to the area law may arise. 
The most studied case is when $A$ is a block of consecutive sites $1,2,\dots, L$ of a one-dimensional critical lattice model. For conformally invariant models, the entanglement entropy scales for large $L$ as   
\be 
S_L=a\frac{c}{6}\ln L+{\rm const}, 
\ee
where $c$ is the central charge of the conformal algebra \cite{holzhey,vidal,Calabrese_Cardy04}, and $a$ is the number of boundary points of the subsystem ($a=1,2$).  

The scaling of the block entanglement entropy has also been studied in the presence of various types of spatial inhomogeneities. Interestingly, the logarithmic scaling holds to be valid in most cases with the difference that $c$ is replaced by some effective central charge $c_{\rm eff}$, which is characteristic of the model and the type of inhomogeneity. 
For the antiferromagnetic XXZ chains with independent, identically distributed (i.i.d.) random couplings, where the strong-disorder renormalization group (SDRG) method \cite{mdh,fisher, fisherxx, im} results in a random singlet phase, the effective central charge was calculated to be $c_{\rm eff}=\ln 2$ \cite{refael, Laflo05, hvlm}.  Furthermore, also  
the distribution  of Schmidt eigenvalues (i.e., the entanglement spectrum)  \cite{fagotti} as well as the full probability distribution of the entanglement entropy across a cut \cite{dmh} were determined.  Similarly, SDRG methods allowed to determine the effective central charge also for models with correlated disorder or inhomogeneities  arising from different types of quasiperiodic modulations \cite{jz, ijz, rsrs}. 
Besides inhomogeneities present overall in the system, also local defects are able to alter the scaling of entanglement entropy, provided they are located at the boundaries of the subsystem.  
For the XXZ chain, a defect coupling results in the saturation of the entanglement entropy ($c_{\rm eff}=0$) in the antiferromagnetic regime, while it is irrelevant in the ferromagnetic regime ($c_{\rm eff}=c=1$) \cite{zpw}. 
In the XX chain, the effect of a defect coupling is marginal, leading to a continuous variation of $c_{\rm eff}$ with the coupling strength \cite{defect, isl09}. 
Similar behavior has been found in the chain of free spinless fermions with a
side-coupled impurity, also known as Fano-Anderson model \cite{eg}.
It is also known for the antiferromagnetic XXZ chains that once the system contains i.i.d. randomness, an additional local defect (i.e., a systematically weaker coupling) at the boundary of the subsystem is irrelevant ($c_{\rm eff}=\ln 2$) \cite{vasseur}.

In models with a local defect, the translational invariance is broken due to the presence of a special site (or coupling). A similar breaking of translational invariance occurs at multiple junctions, i.e., in systems composed of $M$ semi-infinite chains connected to a common central site. Here the local ``topology'' at the central site is different from that of the rest of the system, where it is one-dimensional. Such a junction is known to affect the local critical behavior: For the transverse-field Ising model, the junction induces local ordering for $M>2$, rendering the transition locally discontinuous \cite{itb,ipt,monthus}. Spatial disorder makes the transition continuous for any $M$, with $M$-dependent local order-parameter exponents \cite{stargraph}.  
In such a junction geometry, also the entanglement has been studied recently in a free fermion gas with interactions at the junction \cite{cmv}. In particular, the entanglement entropy of one leg of a multiple junction with the rest of the system has been calculated, and a logarithmic dependence on the length of the leg (assuming constant filling density) has been found with a prefactor showing a universal dependence on the transmission probability for any $M$.   
   
In this work, we aim at going further in exploring the entanglement properties at junction geometries and will study the scaling of entanglement entropy in different disordered models that are accessible by the SDRG method: the random transverse-field Ising model (RTIM), the random XX model, and a fermion lattice gas with random nearest-neighbor hopping. 


\section{Disordered models and the strong-disorder renormalization group}
\label{sec:model}

In this section, we provide a short survey on the SDRG treatment of the transverse-field Ising model, the XX model, and free fermion systems. Based on this, we also derive 
a general upper bound on how much a disordered wire can be entangled with the complementary subsystem.

\subsection{Random transverse-field Ising model}

The first model we will consider is the ferromagnetic transverse-field Ising model defined on an arbitrary lattice by the Hamiltonian 
\be 
H_I=-\sum_{\langle ij\rangle}J_{ij}\sigma_i^x\sigma_j^x - \sum_ih_i\sigma_i^z,
\label{HI}
\ee
where $\sigma_i^{x,z}$ are Pauli matrices on site $i$, the couplings 
$J_{ij}$ and external fields $h_i$ are positive, i.i.d. random variables, and the summation in the first sum goes over nearest neighbors.   
The SDRG procedure for this model is described in details in many works \cite{fisher,im,ki1,ki2}; here, we just recapitulate the essential steps. First, the largest parameter of the model is picked. If it is an external field, $h_i$, site $i$ is decimated and new couplings $\tilde J_{jk}^0$ between all pairs $(j,k)$ of its neighbors are created. According to second-order perturbation calculations, $\tilde J_{jk}^0=J_{ji}J_{ik}/h_i$. 
If there existed a coupling $J_{jk}$ between sites $j$ and $k$ before the decimation, the new renormalized coupling will be $\tilde J_{jk}=J_{jk}+\tilde J_{jk}^0$. A common approximation is the 'maximum rule' according to which  $\tilde J_{jk}=\max\{J_{jk},\tilde J_{jk}^0\}$.
If the largest parameter is a coupling $J_{ij}$ then the spins on site $i$ and $j$ are replaced by a single spin-$1/2$ degree of freedom in an effective transverse field $\tilde h=h_ih_j/J_{ij}$, obtained again by second-order perturbation calculation. If both sites were coupled to a third site $l$ before the fusion, the coupling of the new cluster to $l$ will be $J_{il}+J_{jl}$, while, staying within the maximum rule, it will be $\max\{J_{il},J_{jl}\}$.

As a result of the procedure, the ground state of the system factorizes into a set of independent clusters of (not necessarily adjacent) spins in a ferromagnetically ordered state $\frac{1}{\sqrt{2}}(|\uparrow\uparrow\dots\rangle-|\downarrow\downarrow\dots\rangle)$.
In such a state, the entanglement entropy of a finite subsystem is proportional to the number of clusters containing spins both within as well as outside the subsystem, each such cluster giving a contribution $\ln 2$.

\subsection{Random XX model}

\begin{figure}[t]
\begin{center}
\includegraphics[scale=0.4]{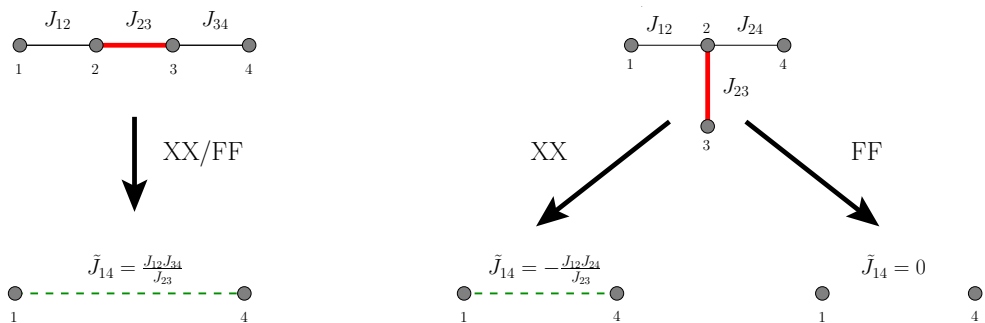} 
\caption{\label{sdrg_steps} SDRG steps for XX models (XX) and free fermions (FF) in the case of two elementary geometries. For the linear configuration the steps for the XX and FF models essentially agree with each other, while for the T shaped geometry they differ.}
\end{center}
\end{figure}

Our second model is the antiferromagnetic XX chain having the Hamiltonian 
\be 
H_{XX}=\sum_{\langle ij\rangle}J_{ij}(S_i^xS_{j}^x+S_i^yS_{j}^y),
\label{hamilton_xx}
\ee
where $S_{i}^{x,y}$ are spin-$\frac{1}{2}$ operators at site $i$. 
The couplings $J_{ij}$ are again i.i.d. random variables, which are initially positive, but, apart from the simple case of a one-dimensional lattice, also negative (i.e. ferromagnetic) couplings may appear during the renormalization. 
In the SDRG procedure, the largest in magnitude coupling $J_{ij}$ is picked, and the pair of sites $(i,j)$, whose state is in zeroth order 
$\frac{1}{\sqrt{2}}(|\uparrow\downarrow\rangle-{\rm sgn}(J_{ij})|\downarrow\uparrow\rangle)$, is eliminated. Assuming that the couplings of site $i$ and $j$ to their neighboring sites are small compared to $J_{ij}$, two types of effective couplings are generated in the second order of the perturbation theory. 
First, each neighbor $k$ of $i$ (except of $j$) is connected with each neighbor  $l$ of $j$ (except of $i$) by a coupling $\tilde J_{kl}^0= J_{ki}J_{jl}/J_{ij}$. 
Second, all pairs $(k,l)$ of neighboring sites of $i$ ($k\neq j$, $l\neq j$) are connected with each other by a coupling $\tilde J_{kl}^0=-J_{ki}J_{li}/J_{ij}$ (and the analogous effective coupling holds also for   all pairs of neighboring sites of $j$). 
If a coupling $J_{kl}$ already existed before the elimination, the perturbative coupling is added to it, $\tilde J_{kl}=J_{kl}+\tilde J_{kl}^0$, while using the maximum rule, $\tilde J_{kl}=\max\{J_{kl},\tilde J_{kl}^0\}$. The SDRG steps for two elementary geometries are depicted in Fig.~\ref{sdrg_steps}.

This renormalization procedure, which is asymptotically exact at least in one dimension \cite{fisherxx}, results in a factorized ground state comprising pairs of spins in fully entangled states $\frac{1}{\sqrt{2}}(|\uparrow\downarrow\rangle\pm|\downarrow\uparrow\rangle)$. The entanglement entropy of a subsystem is thus given by the number of pairs with precisely one constituent spin in the subsystem, multiplied by $\ln 2$.

\subsection{Free-fermion model with off-diagonal disorder}

Finally, we will also consider free-fermion models with random nearest-neighbor hopping terms. The corresponding Hamiltonian is
\be 
H_{FF}=-\sum_{\langle ij\rangle}J_{ij}(c_i^{\dagger}c_j+c_j^{\dagger}c_i), 
\label{hamilton_ff}
\ee
where $c_i^{\dagger}$ and $c_i$ are fermion creation and annihilation operators on site $i$, respectively, while the transition amplitudes $J_{ij}$ are positive, i.i.d random variables.  
For a one-dimensional chain, the Hamiltonian of the XX model in Eq.~(\ref{hamilton_xx}) can be mapped to the free fermion Hamiltonian in Eq. (\ref{hamilton_ff}) by the well-known Jordan-Wigner transformation. Consequently, the SDRG rules in one-dimension are identical to those of the XX chain.  However, an SDRG scheme for the free-fermion model on a general lattice has, to our knowledge, not been formulated so far. 
 
In order to derive the general SDRG scheme, one needs to consider only two types of basic four-site graphs, a linear and a T shaped graph depicted in Fig.~\ref{sdrg_steps}, where
 the largest coupling is between sites 2 and 3 and  the rest of the terms in the Hamiltonian is treated as perturbation. 

 This means that the unperturbed Hamiltonian is the same for both graphs, $H_0= - J_{23}(c_2^{\dagger}c_3+c_3^{\dagger}c_2)$, which has a  four-fold degenerate ground-state space.
Let us introduce the following notation for the basis vectors spanning the ground-state subspace
\begin{equation} \label{eq:SDRG_basis}
  \begin{array}{l} 
| 00 \rangle =  \frac{1}{\sqrt{2}}(c_2^{\dagger} + c_3^{\dagger}) | 0000 \rangle_F, \\
| 10 \rangle = c^\dagger_1 | 00 \rangle,\\
| 01 \rangle = c^\dagger_4 | 00 \rangle,\\
| 11 \rangle = c^\dagger_1  c^\dagger_4 | 00 \rangle,
\end{array}
\end{equation}
where $| 0000 \rangle_F$ is the Fock vacuum for the four-site block.  
For the linear graph case depicted in the left part of Fig.~\ref{sdrg_steps}, the perturbation Hamiltonian is given by $H^{(a)}_p= - J_{12}c_1^{\dagger}c_2-   J_{34}c_3^{\dagger}c_4 + h.c. $, while for T shaped one, it is  $H^{(b)}_p= - J_{12}c_1^{\dagger}c_2-   J_{24}c_2^{\dagger}c_4 + h.c. $. Using degenerate perturbation theory up to second-order, we can write the effective Hamiltonian acting on the unperturbed ground-state subspace as
\be
H^{(v)}_{\rm eff}= P H_p^{(v)} P +  P H_p^{(v)} \frac{1-P}{E-H_0} H^{(v)}_p P, \; \; \; \quad v=a,b,
\ee
where  $E$ is the ground-state energy and $P$ is the projection onto the ground-state subspace. Using this expression and the basis given by Eq.~\ref{eq:SDRG_basis}, we obtain that for the linear graph geometry, the effective Hamiltonian has the following form
\begin{equation*}
  \begin{array}{l}  \nonumber
\langle 00| H^{(a)}_{\rm eff} | 00\rangle = \langle 10| H^{(a)}_{\rm eff} | 10\rangle =\langle 01| H^{(a)}_{\rm eff} | 01\rangle =\langle 11| H^{(a)}_{\rm eff} | 11\rangle= - \frac{J^2_{12} + J^2_{34}}{2 J_{23}}\\
\langle 10| H^{(a)}_{\rm eff} | 01\rangle = \langle 01| H^{(a)}_{\rm eff} | 10 \rangle = - \frac{J_{12}J_{34}}{J_{23}}, \nonumber
\end{array} \nonumber
\end{equation*}
with all other matrix elements being zero. This means that  the effective Hamiltonian can be written as
\be
 H^{(a)}_{\rm eff}=   - \frac{J_{12}J_{34}}{J_{23}} (c_1^{\dagger}c_4+c_4^{\dagger}c_1) - \frac{J^2_{12} + J^2_{34}}{2 J_{23}} \, \mathbf{1},
\ee
which, equivalently to the SDRG result for the XX chain, provides an effective hopping amplitude  $\tilde J_{14}=  \frac{J_{12}J_{34}}{J_{23}} $ between sites $1$ and $4$ (and a constant term).
For the T shaped geometry depicted on the right part of Fig.~\ref{sdrg_steps},  the second order perturbation theory 
yields 
\begin{equation*}
  \begin{array}{l}  \nonumber
\langle 00| H^{(b)}_{\rm eff} | 00\rangle = \langle 10| H^{(b)}_{\rm eff} | 10\rangle =\langle 01| H^{(b)}_{\rm eff} | 01\rangle =\langle 11| H^{(b)}_{\rm eff} | 11\rangle= - \frac{J^2_{12} + J^2_{34}}{2 J_{23}}\\
\langle 10| H^{(b)}_{\rm eff} | 01\rangle = \langle 01| H^{(b)}_{\rm eff} | 10 \rangle = 0, \nonumber
\end{array} \nonumber
\end{equation*}
with also the other matrix elements being zero. This, unlike tehe XX case, simply yields a constant effective Hamiltonian $H^{(b)}= - \frac{J^2_{12} + J^2_{34}}{2 J_{23}} \, \mathbf{1} $, meaning that the effective transition amplitude is zero at this order of the perturbation, i.e. $\tilde J_{14}=0$.

This implies the  following SDRG scheme for a generic graph with a  random free-fermion model.
First, the block of sites having the largest transition amplitude $J_{ij}$ is selected. This block, which has the ground state $\frac{1}{\sqrt{2}}(|10\rangle+|01\rangle)$ at zeroth order, is eliminated. Then, as for the XX model, each neighboring site $k$ of $i$ (except of $j$) is connected with each neighbor $l$ of site $j$ (except of $i$) by a coupling obtained in the second order of perturbation theory as $\tilde J_{kl}^0=J_{ki}J_{jl}/J_{ij}$. 
The difference compared to the SDRG rules of the XX model is that the couplings between neighboring sites of site $i$ (and $j$) are zero to second order, so the above scheme is complete. 
In case a new coupling is generated and the two sites were already connected before, we proceed as described for the previous two models.  
The resulting ground state obtained by the SDRG procedure is a product of one-particle states $\frac{1}{\sqrt{2}}(|10\rangle+|01\rangle)$ of blocks of two sites. The entanglement entropy of a finite subsystem is given by the number of blocks with precisely one site inside of the subsystem, multiplied by $\ln 2$.

\subsection{Junction geometries and the SDRG}

\begin{figure}[t]
\begin{center}
\includegraphics[scale=0.24]{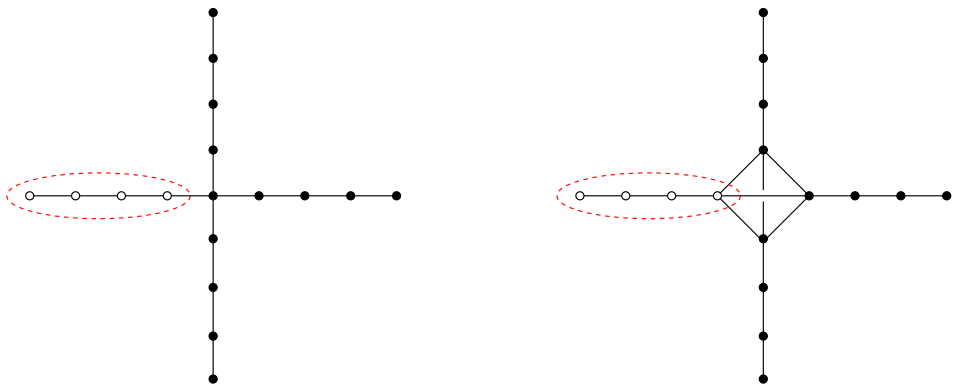}
\caption{\label{juncgeo} The star junction and the completely connected junction geometries for $M=4$ arms. The entanglement between one arm (encircled by the dashed red line) and the rest of system is studied in the paper.}
\end{center}
\end{figure}

Here, we consider the above three models on a lattice, which is composed of $M$ one-dimensional chains with $L$ sites (called \emph{arms}), the first sites of which are connected with each other by couplings $J^0_{mn}$. Here, we use the indices  $m,n=1,2,\dots M$ to label the arms in this junction geometry.  
In the case when all couplings $J^0_{mn}$ are non-zero and are drawn independently from the same distribution as that of the bulk couplings, we shall speak of a fully connected junction. The junction can also be partially connected, by which is meant that the lattice is still connected, but some of the arms are not coupled with each other, i.e.  $J^0_{mn}=0$ for some $(m,n)$, while the rest of the couplings are i.i.d. variables like in the interior of the arms.
A special case of partially connected junctions is when all couplings are zero except of $J_{1n}^0$, $n=2,3,\dots M$, which we call a \emph{star junction}. 
In all cases, we are interested in the average entanglement entropy of one arm of the lattice with the rest of the system (for the star junction case, we will not consider the central site to be part of the arm), see Fig.~\ref{juncgeo}  depicting the cases of  the star and the fully connected junction.
Based on that the subsystem is still one-dimensional, furthermore knowing the results about the homogeneous fermion gas at junction geometries \cite{cmv}, we expect a logarithmic asymptotical scaling of the average entanglement entropy with the size of the subsystem $L$, as 
\be 
S(L)=\frac{c_{\rm eff}}{6}\ln L + {\rm const}, 
\label{SL}
\ee 
with some effective central charges which may depend on the details of the junction geometry. 

It is well known that the one-dimensional variants of the above models remain one-dimensional under the action of the SDRG transformation, moreover, their local parameters (couplings and fields) remain independent. Higher dimensional lattices are no longer invariant due to the proliferation of couplings between farther sites, the underlying lattice becoming gradually more and more connected; in addition to this, the parameters on different positions become correlated with each other.  The question is what happens to the topology at a junction of $M$ one-dimensional chains during the SDRG procedure. 

For the RTIM on a star junction of chains, the first elimination (field decimation) of the central site transforms the junction to a fully connected one. 
At any stage of the renormalization, the lattice will consist of a central full graph of size $1\le \mathcal{M}\le M$, each site $i$ of which holds $N_i$ arms ($\sum_{i=1}^{\mathcal{M}}N_i=M$). These geometries transform into each other either by a field decimation on site $i$ of the full graph, which increases the size $\mathcal{M}$ of the full graph by $N_i-1$ or by a coupling decimation within the full graph, which reduces $\mathcal{M}$ by $1$. 
It is not difficult to see that any partially connected junction becomes trapped in this absorbing set of possible geometries during the SDRG procedure after a transient period. 
Therefore, the large scale properties of system, such as the prefactor $c_{\rm eff}$ in the expected scaling form of the entanglement entropy in Eq.~(\ref{SL}) will be the same irrespective of the type of junction; it can influence only the additive constant in Eq.~(\ref{SL}). 

For the XX model, the situation is simpler than for the RTIM. Here, any type of initial junction will tend toward the fully connected one owing to the proliferation of couplings at the junction. Once the full connectedness is reached, it will remain unchanged, so the fully connected one is an invariant junction geometry for this model. The effective central charge is therefore again insensitive with respect to the type of junction.  

The behavior of the junction geometry under the SDRG procedure in the hopping model is different from that of the previous two models, as there are certain constraints which limit the formation of full connectedness. To see this, let us assume that the hopping model is defined on a bipartite lattice. This means that it is composed of two sublattices, $A$ and $B$, such that all sites of $A$ are connected exclusively with sites of $B$, and vice versa. In this case, the model has a sublattice symmetry, which implies that the energy spectrum is symmetric around zero. When a pair of sites is decimated in the SDRG procedure, the sites must be on different sublattices, and the generated new couplings connect exclusively sites on different sublattices, thus the SDRG procedure leaves the lattice bipartite. This also means that the SDRG approximation up to second order conserves the sublattice symmetry of the model. 
What is the consequence of this property of the SDRG at junction geometries? 
If the arms are connected in such a way that the lattice is not bipartite, then the junction will reach the full connectedness during the SDRG procedure. If, however, the arms are coupled in such a way that the lattice is bipartite and 
the end sites of $1\le m<M$ arms are on one sublattice while the end site of the other $M-m$ arms on the other sublattice, then couplings between arms on the same sublattice will never form, but the two sublattices will become fully interconnected at late stages of the renormalization. Once full interconnectedness is reached, the effect of a decimation of coupling $J_{ij}$ within the central part is that the arms attached to site $i$ and $j$ will be connected through their new end sites to the other sublattice. 
The invariant junction geometry thus depends on whether the initial lattice is bipartite or not, and in the former case, on the number of arms on the same sublattice. For $M$ odd, the number of different possible invariant geometries is therefore $(M+1)/2$ while, for $M$ even ($M>2$), it is $M/2+1$. These different classes of initial junction geometries are expected to be characterized by different effective central charges.

\subsection{Entanglement entropy capacity of a random chain}

Before presenting numerical results on the entanglement entropy scaling, we shall establish upper bounds on the average entanglement entropy of a chain of length $L$ for all the three models under study, which hold universally, irrespective of the other subsystem to which the chain is coupled. This is based on the fact that, in these models, there is only a vanishing fraction of degrees of freedom in finite subsystems, which are able to contribute to the entanglement entropy (within the SDRG approximation). 
 
To see this, let us first consider the random XX chain and a triplet of adjacent couplings $J_{12},J_{23},J_{34}$ in it, for which 
\be
J_{23}>J_{12},J_{34}.
\label{cond}
\ee
In a decimation step, the generated new coupling never exceeds the decimated ones, therefore the lateral couplings $J_{12}$ and $J_{34}$ may decrease during the SDRG due to decimations of their neighboring couplings but they will never be greater than $J_{23}$. 
Thus, at some stage of the SDRG, $J_{23}$ is decimated, meaning that spin $2$ and $3$ form a singlet. Consequently, the triplets of couplings fulfilling the condition in Eq.~(\ref{cond}), called 'local maxima' can be decimated (in an arbitrary order) even if $J_{23}$ is not the global maximum, and the resulting ground state will be the same as obtained by the traditional implementation of the SDRG. 
The condition for the validity of such a 'local' SDRG is that the newly generated couplings are not greater than the decimated ones. Apart from the one-dimensional lattice, this cannot be guaranteed due to the addition of couplings, see the previous subsections. Nevertheless, applying the maximum rule, this condition is fulfilled and the local SDRG is applicable \cite{ki1,ki2}. 

Let us now assume that the subsystem is an XX chain with $1,2,\dots,L$ sites, the first site of which is coupled to a spin (labeled by $0$) of the other subsystem (the environment), about which we have no further informations, by the coupling $J_{01}$. The spin pairs at the local maxima of the sequence of couplings can surely be arranged into singlets, but some spins may remain for which we cannot decide, without any informations on the environment, which spin to pair up with. For instance, if $J_{01}>J_{12}$, spin $1$ would form a singlet with spin $0$ of the environment if the couplings of $0$ to the other spins of the environment were smaller than $J_{01}$; otherwise it can happen that spin $0$ couples to another spin of the environment, leaving behind a weak ($<J_{12}$) renormalized coupling between  the environment and spin $1$, so that spin $1$ ultimately forms a singlet with spin $2$.    
Formally, the local SDRG procedure ends up with a monotonically decreasing sequence of effective couplings between 'undetermined' spins, whose state is not exclusively determined by the parameters of the subsystem but also by those of the particular environment. Only these spins are able to be entangled with the environment, so knowing their number $N_u(L)$, we can obtain an upper bound on the entanglement entropy, which we call entanglement capacity. For a given $N_u$, the highest possible  entanglement entropy is obtained if all the $N_u$ spins are separately maximally entangled with the environment, for which case $S=N_u\ln 2$. 
It is straightforward to show that, for any subsystem with a decreasing sequence of couplings, one-dimensional random XX chains exist as environments, with which the entanglement entropy of the subsystem is maximal.  

In Ref. \cite{jkri}, it was argued by using a relationship between the sequence of couplings and random walks that the average number of undetermined spins is proportional to $\ln L$ but the proportionality constant was left unknown. Here, we go further and determine the proportionality constant by a different approach relying on the SDRG method itself, which goes as follows.

Let us start with a semi infinite chain with spins $0,1,2,\dots$, among which sites $1,2,\dots$ constitute the subsystem. Apply now the SDRG procedure in its original formulation, in which the global energy scale $\Omega=\sup_i{J_i}$ is gradually reduced by singlet decimations, with the modification that the leftmost coupling is not allowed to be decimated. Until the energy scale $\Omega$ is higher than the leftmost coupling $\tilde J_0$ (which may have been reduced compared to its initial value due to decimations on its right), the renormalization goes in the usual way. At some stage, however, the energy scale $\Omega$ will reach $\tilde J_0$, meaning that the leftmost coupling is ready for decimation. Since $\Omega=\tilde J_0$, the coupling next to $\tilde J_0$ must be smaller than $\tilde J_0$, so the spin next to spin $0$ will be an undetermined spin. 
At this point, we delete the leftmost spin ($0$) and the coupling $\tilde J_0$, so that the undetermined spin just found will be the leftmost spin of the lattice.
Continuing the SDRG procedure down to the scale where again the leftmost coupling is picked, we can identify the spin next to the leftmost one as the second undetermined spin. The leftmost spin and its coupling are deleted again, and by repeating this procedure, we find a new undetermined spin each time the actually leftmost coupling is ready for decimation.    
The SDRG procedure for the random XX chain is analytically tractable, for the details, see Ref.~\cite{fisherxx}. The distribution of reduced couplings $\beta_i=\ln(\Omega/J_i)$ is attracted by the fixed-point distribution
\be
P_{\Gamma}(\beta)=\frac{1}{\Gamma}e^{-\beta/\Gamma},
\label{P}
\ee
at low energies, 
where $\Gamma=\ln(\Omega_0/\Omega)$ and $\Omega_0$ depends on the initial distribution of couplings.
Furthermore, the couplings on different sites are independent, and for a semi-infinite chain, the end coupling follows the same fixed-point distribution as the bulk ones \cite{fisher}. As a consequence, the actual end coupling in the modified SDRG procedure has the fixed-point distribution in Eq.~(\ref{P}) at any (late) stage. When the logarithmic energy scale increases infinitesimally from $\Gamma$ to $\Gamma+d\Gamma$, the probability of decimating the end coupling or, equivalently, of finding an undetermined spin is therefore $P_{\Gamma}(0)d\Gamma$.  
Thus, the average number of undetermined spins found when the renormalization is carried out up to some large scale $\Gamma$ increases asymptotically as 
\be 
N_u(\Gamma)=\int^{\Gamma}P_{\Gamma'}(0)d\Gamma'=\ln\Gamma + {\rm const.}
\ee
Using that the energy and length scale are related as  $\Gamma_L\sim \sqrt{L}$ \cite{fisherxx}, we obtain for the finite-size scaling of the entanglement capacity of the random XX chain: 
\be 
S_u(L)=\frac{\ln 2}{2}\ln L + {\rm const.}
\ee
for large $L$. 

For the random hopping model, the SDRG scheme in one-dimension is identical to that of the XX chain, so we obtain the same result for the entanglement entropy capacity. 
For the transverse-field Ising chain, arranging the couplings and fields in an alternating fashion, $h_1,J_1,h_2,J_2,\dots$, the SDRG procedure is formally identical to that of the XX chain. For a chain with $L$ spins, the length of the sequence of parameters is therefore $2L$, from which the modified SDRG leaves on average a sequence of length $\frac{1}{2}\ln(2L)+{\rm const}$. The number of undetermined spins is the half of it, so the entanglement entropy capacity\footnote{Here by entanglement entropy capacity we mean the maximal entanglement a disordered wire can share with the rest of a system which is joined to it. This is a type of integrated version of the entanglement susceptibility introduced in \cite{zv}.} of the random transverse-field Ising chain scales asymptotically as 
\be 
S_u(L)=\frac{\ln 2}{4}\ln L + {\rm const.}
\ee
Using the maximum rule approximation, which is asymptotically exact at infinite-disorder fixed points \cite{im}, for junction geometries, the local SDRG method is applicable, consequently the above upper bounds are valid. 
According to these results, in any junction geometry, we have the upper bound for the effective central charge
\be 
c_{\rm eff}\le 3\ln 2
\ee
for the XX and the hopping model, while 
\be 
c_{\rm eff}\le \frac{3}{2}\ln 2
\ee
for the RTIM.

\section{Numerical results} \label{sec:num_res}

We continue by implementing numerically the SDRG procedure for the three systems under study, and obtain estimates for the corresponding effective central charges. For the free fermion model the SDRG results are tested against exact numerical diagonalization, and an excellent agreement is found. In the case of the XX model, we conjeture a formula for the effective central charge in terms of the number of arms, which we conjecture to be asymptotically exact in the large $M$ limit.

\subsection{Transverse-field Ising model}

We have implemented the SDRG method numerically for all three models at junction geometries and calculated the average entanglement entropy. 
For the RTIM, where the type of junction geometry is irrelevant, we used star junctions with $M=2,3,4,5,6$ arms and with different arm lengths $L=2^k$, $k=3,4,\dots,13$. 
The couplings and transverse fields were drawn independently from a uniform distribution with the support $(0,1)$. Such a distribution fulfills the criticality condition in the bulk of the system, $\overline{\ln J}=\overline{\ln h}$, where the overbar denotes averaging \cite{fisher}.  
We have mainly used the maximum rule, which allows for the application of the local SDRG method. This can be implemented in an efficient way by first renormalizing the arms independently to a length $\sim\ln L$, which can be carried out by $O(ML)$ operations, then applying the SDRG to this pre-renormalized (much smaller) system.
For all $M$ and $L$, the entanglement entropy of one arm was averaged over $10^8$ independent random samples.
In order to check the validity of this approximation, we also applied the SDRG with the sum rule, which requires a much longer computational time, and for which the number of samples was at least $10^6$. 

\begin{figure}[t]
\begin{center}
\includegraphics[width=7.7cm]{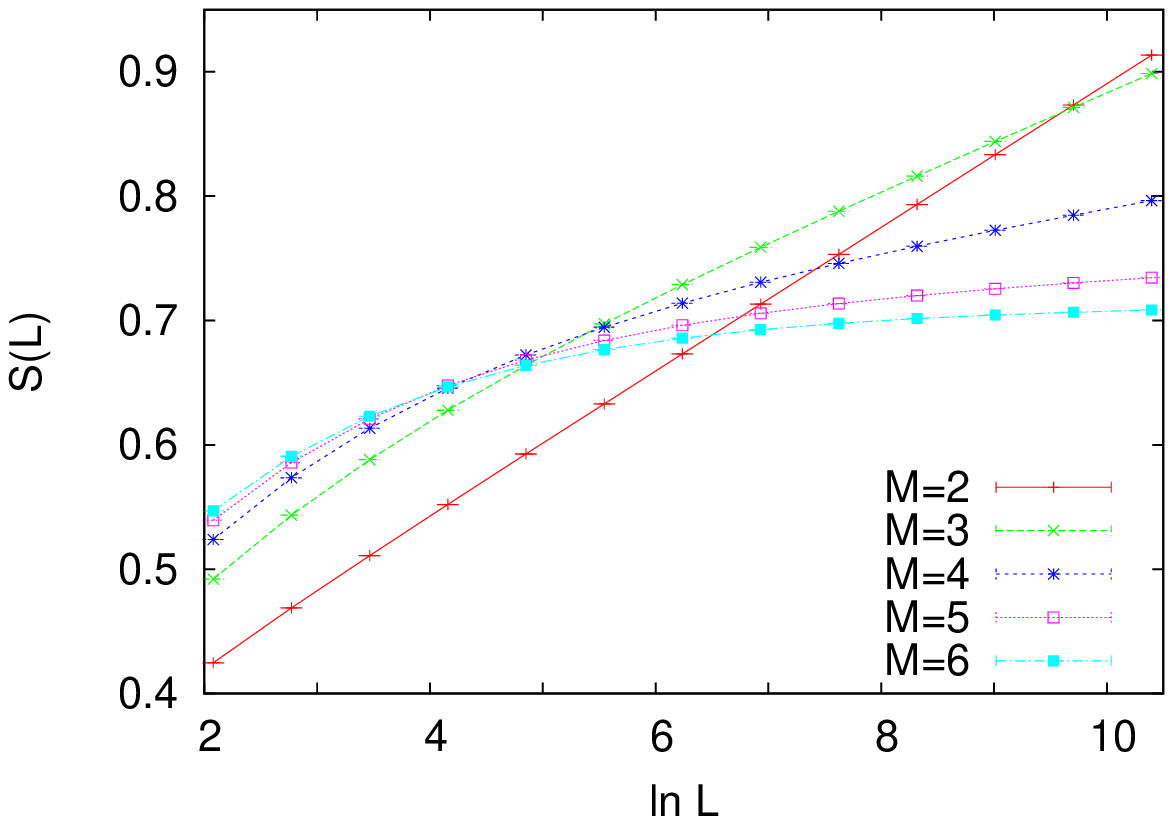}
\includegraphics[width=7.7cm]{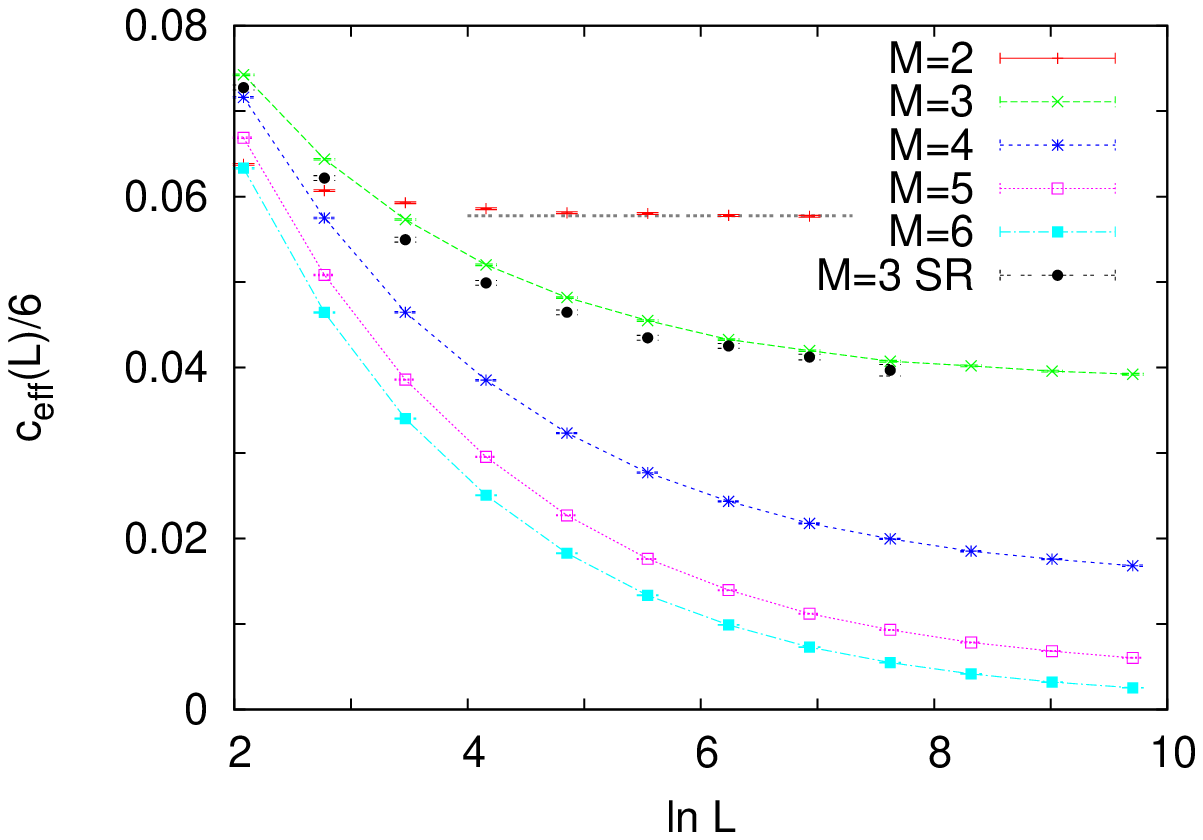} 
\caption{\label{rtim} Left. The dependence of the average entanglement entropy on the size of the subsystem in the RTIM at star junctions for different values of $M$.    
Right. The size-dependence of the two-point fits for the effective central charge. The data denoted by SR were obtained by the SDRG method with the sum rule; all other data were obtained by using the maximum-rule approximation.   
}
\end{center}
\end{figure}

The average entanglement entropy $S(L)$, as well as the two-point fits of the effective central charge $c_{\rm eff}(L)=6[S(2L)-S(L)]$ as a function of the subsystem size $L$ are shown in Fig. \ref{rtim}. 
As can be seen, the two-point fits $c_{\rm eff}(L)$ for a given $M$ tend very slowly to their asymptotic value. Nevertheless, we can read off from the finite-$L$ data that the $M$-dependent asymptotic values must decrease with increasing $M$, and presumably tend to zero in the large-$M$ limit. 

\subsection{XX model}

For the random XX model, we studied numerically fully connected junctions, which are the only invariant junction geometries. 
The couplings were drawn independently from a uniform distribution in the interval $(0,1)$. 
We applied the maximum-rule approximation, which allows us to use the local SDRG. Similarly to the RTIM, each arm (of length $2^k$, $k=3,4,\dots,13$) was pre-renormalized and coupled to a site of a fully connected part of the system, so that the size of the subsystem was $2^k+1$. This means, that for $M$ odd, the total number of sites is odd, so the SDRG leaves one spin unpaired. Nevertheless, this does not influence the asymptotic value of the effective central charge.   
Note that, although during the SDRG also negative couplings are generated, using the maximum-rule approximation, it is sufficient to record the magnitude of couplings. 
The number of arms went up to $M=40$, while the number of independent random samples was $10^8$ for $M\le 8$, and $10^7$ otherwise. In each sample, the entanglement entropy of every arm with the rest of the system was calculated and added to the average. 

\begin{figure}[t]
\begin{center}
\includegraphics[width=7.7cm]{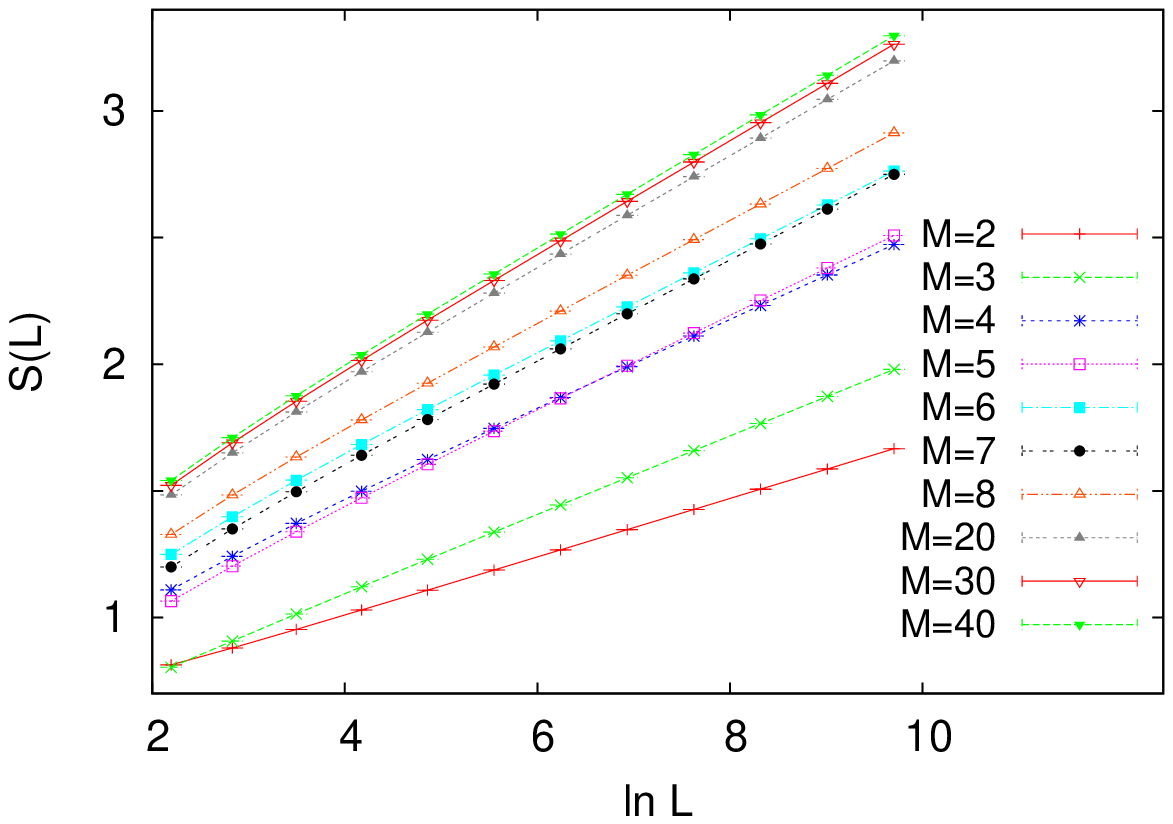}
\includegraphics[width=7.7cm]{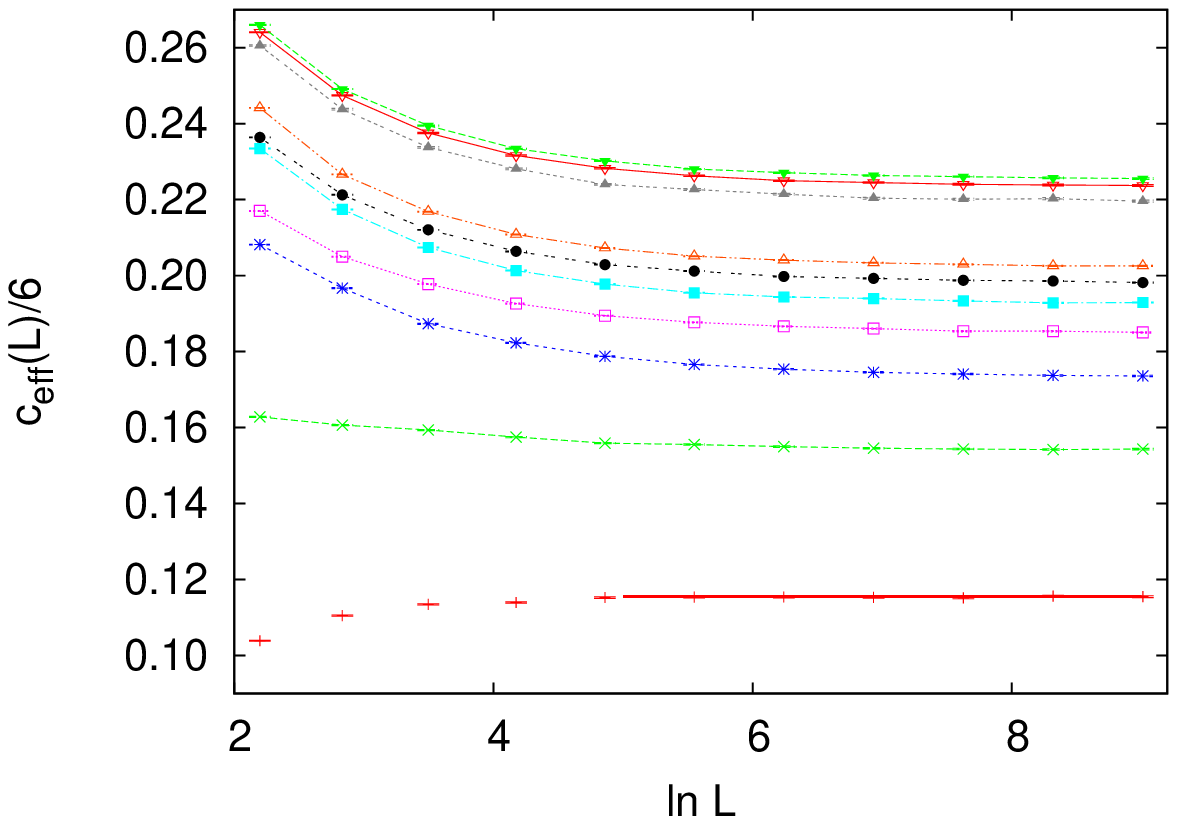} 
\caption{\label{xx} Left. The dependence of the average entanglement entropy on the size of the subsystem in the random XX model at fully connected junctions for different values of $M$.    
Right. The size-dependence of the two-point fits for the effective central charge. All data were obtained by the SDRG method with the maximum-rule approximation.   
}
\end{center}
\end{figure}

\begin{figure}[t]
\begin{center}
\includegraphics[width=7.7cm]{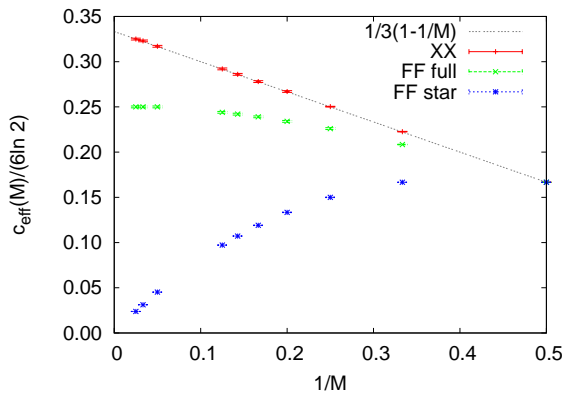}
\caption{\label{ceff} Numerical SDRG estimates of $c_{\rm eff}/(6\ln 2)$ for the XX model and for the free-fermion model with fully connected and star junction plotted against $1/M$.  
}
\end{center}
\end{figure}

The dependence of the average entanglement entropy as well as that of the corresponding two-point fits of the effective central charge on $L$ are shown in Fig. \ref{xx} for different values of $M$. 
The relatively fast convergence of $c_{\rm eff}(L)$ with increasing $L$ allows us to give estimates on the asymptotical value $c_{\rm eff}$ for each $M$, which can be found in Table. \ref{table_I} and are plotted against $1/M$ in Fig. \ref{ceff}. 
\begin{table}[h]
\begin{center}
\begin{tabular}{|c|l|l|l|}
\hline  $M$  &  XX model  & free-fermion (full) & free-fermion (star) \\ \hline
2            &  0.1666(3) & 0.1666(3)           & 0.1666(3)    \\ \hline
3            &  0.2226(3) & 0.2084(3)           & 0.1667(3)    \\ \hline
4            &  0.2503(4) & 0.226(1)            & 0.1500(3)    \\ \hline
5            &  0.267(1)  & 0.234(1)            & 0.1334(3)    \\ \hline
6            &  0.278(1)  & 0.239(1)            & 0.1191(3)    \\ \hline
7            &  0.286(1)  & 0.242(1)            & 0.1072(3)    \\ \hline
8            &  0.292(1)  & 0.244(1)            & 0.0972(2)    \\ \hline
20           &  0.317(1)  & 0.250(1)            & 0.0452(2)    \\ \hline
30           &  0.323(1)  & 0.250(1)            & 0.0312(2)    \\ \hline
40           &  0.325(1)  & 0.250(1)            & 0.0238(1)    \\ \hline
\end{tabular}
\end{center}
\caption{\label{table_I} Numerical SDRG estimates of $c_{\rm eff}/(6\ln 2)$ for the XX model and for the free-fermion model with fully connected and star junction, for different number $M$ of arms.}
\end{table}
Here, as opposed to the RTIM, the effective central charge increases with the number $M$ of arms. 
It is remarkable that the numerical data fit well to the form linear in $1/M$: 
\be \label{XX_cent_charge}
c_{\rm eff}(M)=2\ln 2\left(1-\frac{1}{M}\right). 
\ee
Note that this form is exactly valid for $M=2$, and gives zero in the trivial case $M=1$. 
In the large-$M$ limit, $c_{\rm eff}(M)$ seems to converge to a limit, which is well below the upper bound $3\ln 2$ found in the previous section. 

\subsection{Free-fermion model}

For the free-fermion model, where the type of junction geometry is of importance with respect to the effective central charge, we studied two cases: the star junction and the fully connected junction. 
The transition amplitudes were drawn from a uniform distribution in the interval $(0,1)$. In both cases, we applied a pre-renormalization of the arms; in the fully connected case, the maximum-rule approximation was used. 
A peculiarity of the hopping model at a star junction is that, similar to the one-dimensional model, couplings are never added, so it is not needed to resort to the maximum rule.  
The number of random samples was $10^8$ for the star junction, and at least $10^7$ for the fully connected one. In each sample, the entanglement entropy between every arm and the rest of the system was calculated.

\begin{figure}[t]
\begin{center}
\includegraphics[width=7.7cm]{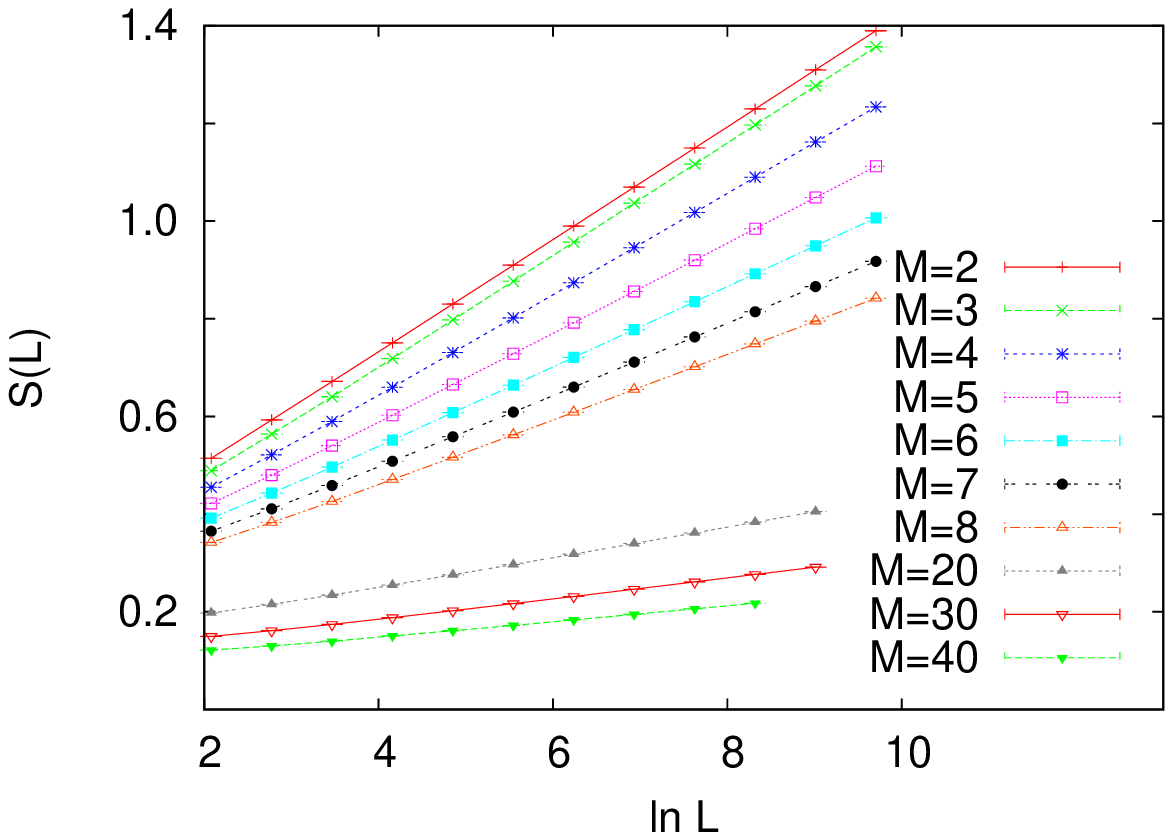}
\includegraphics[width=7.7cm]{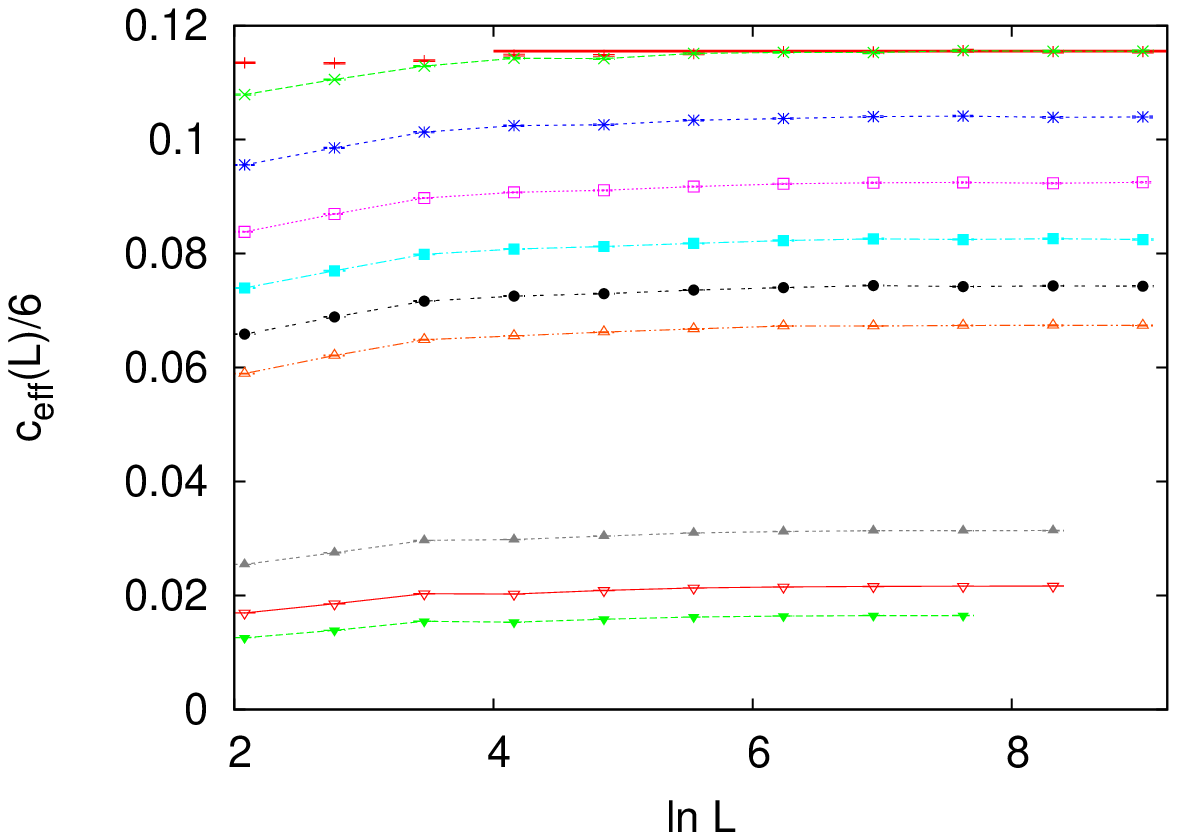} 
\caption{\label{fstar} Left. The dependence of the average entanglement entropy on the size of the subsystem in the random free-fermion model at star junctions for different values of $M$. The data were obtained by the SDRG method.   
Right. The size-dependence of the two-point fits for the effective central charge.}
\end{center}
\end{figure}

Numerical results obtained for the star junction for different values of $M$ are shown in Fig. \ref{fstar}. Estimates of the asymptotic values of $c_{\rm eff}$ are summarized in Table \ref{table_I} and are plotted as a function of $1/M$ in Fig. \ref{ceff}.  
As can be seen, the effective central charge for $M=3$ coincides with that of the one-dimensional chain ($M=2$), $c_{\rm eff}=\ln 2$ within the error of estimation. For $M>3$, the effective central charge decreases monotonically with $M$, and presumably tends to zero in the limit $M\to\infty$. 

\begin{figure}[t]
\begin{center}
\includegraphics[width=7.7cm]{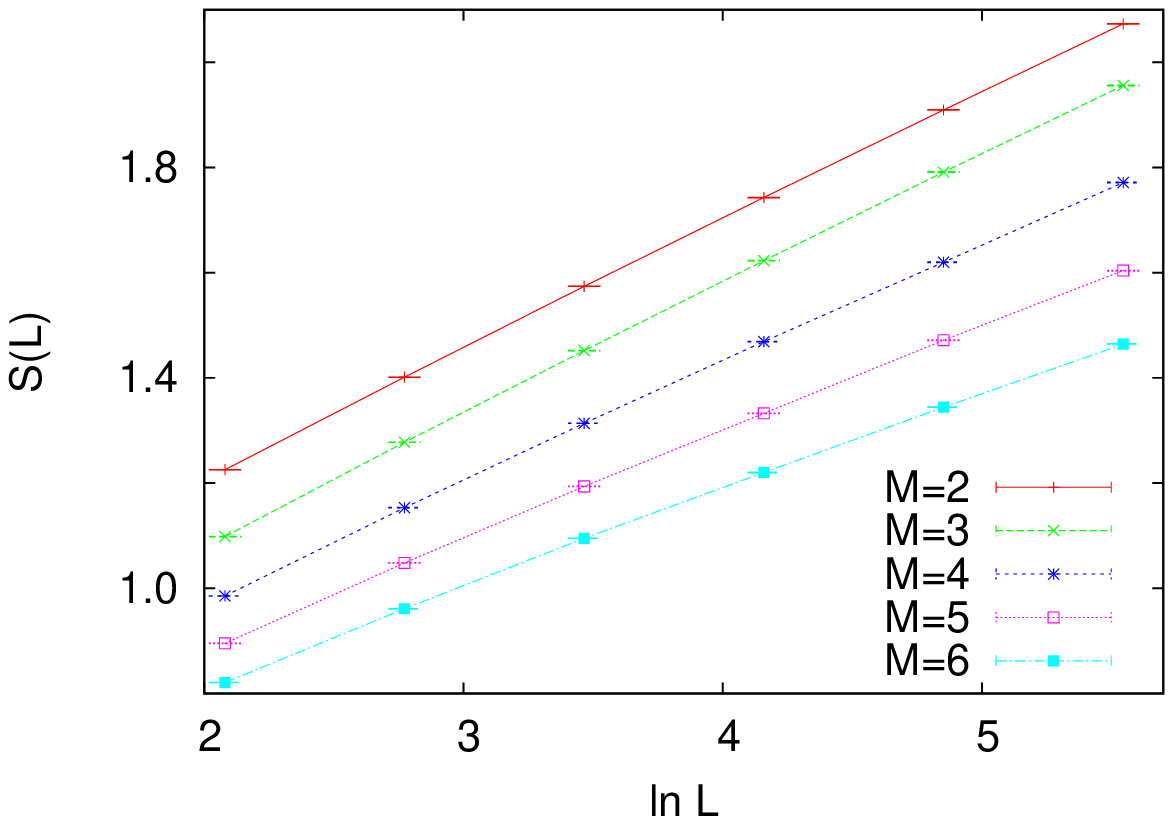}
\includegraphics[width=7.7cm]{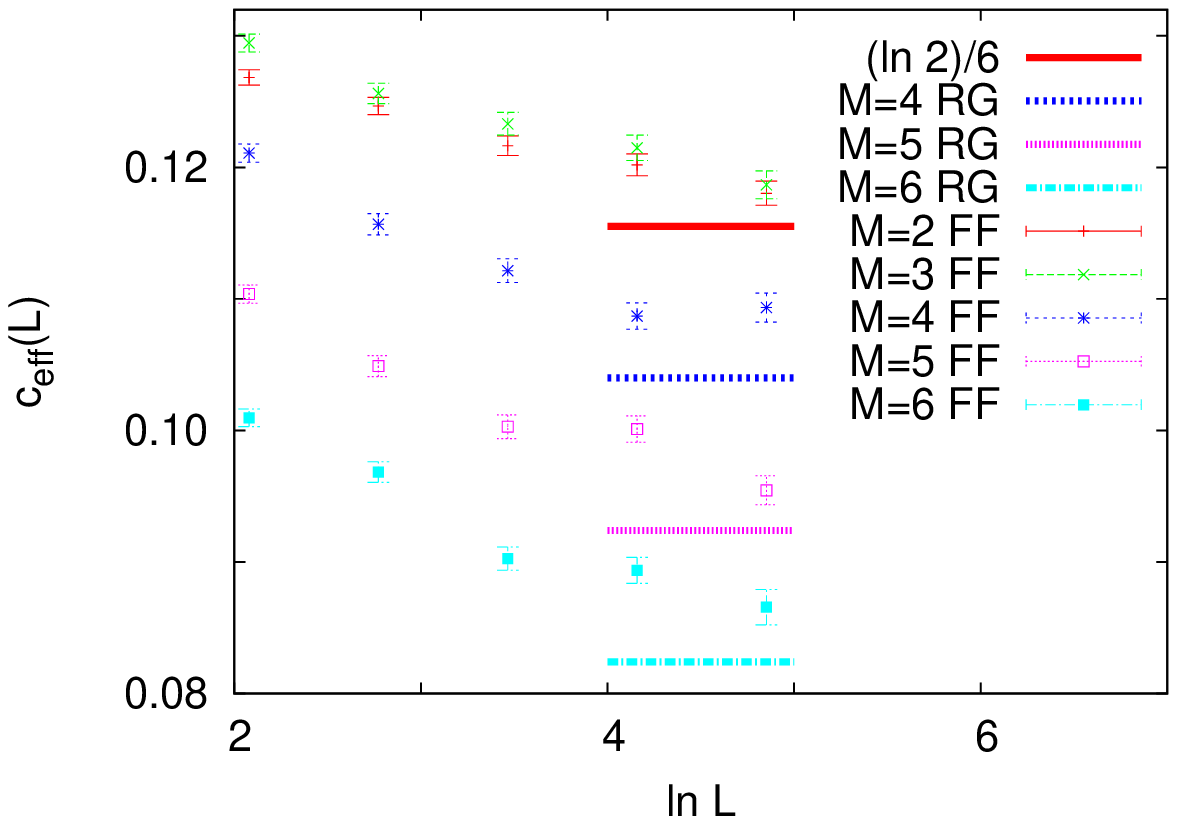}
\caption{\label{diag} Left. The dependence of the average entanglement entropy on the size of the subsystem in the random free-fermion model at star junctions for different values of $M$. The data were obtained by exact diagonalization.    
Right. The size-dependence of the two-point fits for the effective central charge. The horizontal lines indicate the estimates obtained by the SDRG method.}
\end{center}
\end{figure}

We calculated the entanglement entropy also by exact diagonalization \cite{vidal, peschel03}. Here, we considered a double junction geometry, i.e. not only the first site of arms were coupled to an extra site, but also site $L$ of the arms were coupled to another extra site. The system sizes available by this method, which is limited by the occurrence of numerical instabilities in certain samples, are much smaller than for the SDRG method. The number of random samples was $10^6$ for each size. As can be seen in Fig. \ref{diag}, the asymptotic regime has not reached with the sizes available by the method; nevertheless the two-point fits $c_{\rm eff}(L)$ seem to tend toward the estimates obtained by the SDRG method.  

\begin{figure}[t]
\begin{center}
\includegraphics[width=7.7cm]{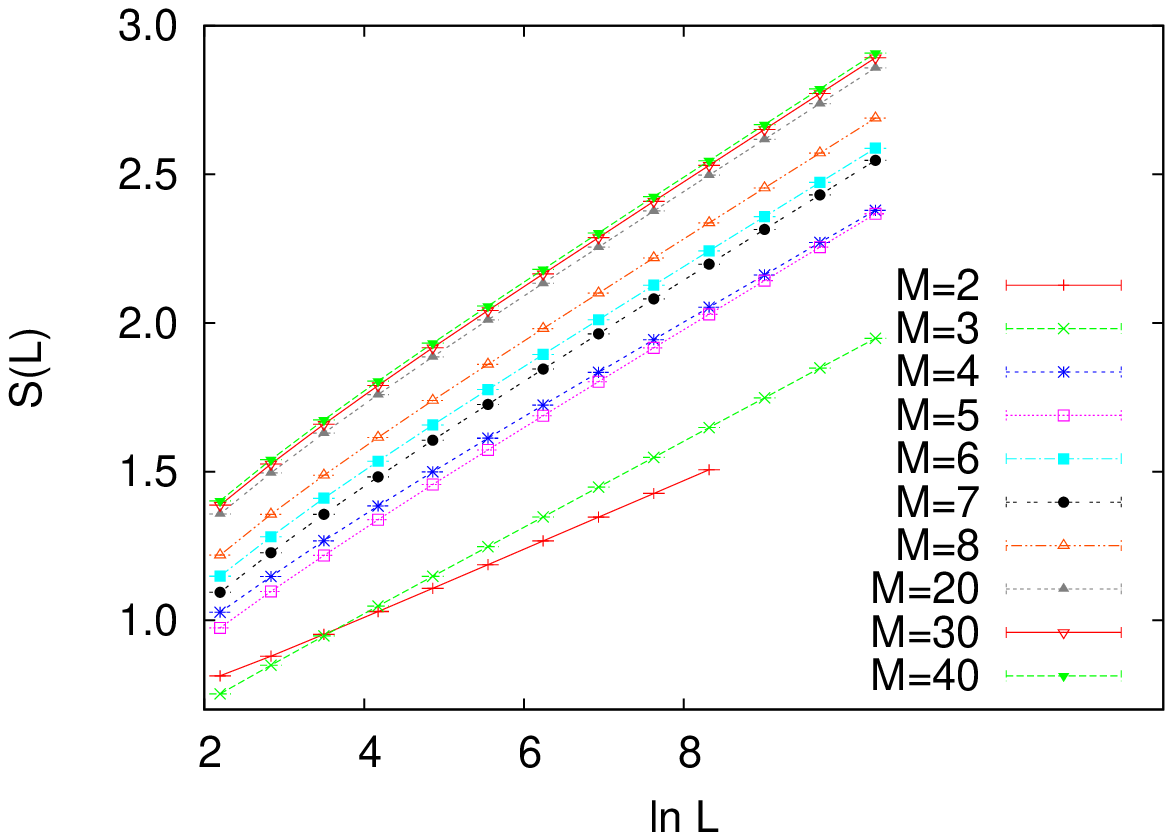}
\includegraphics[width=7.7cm]{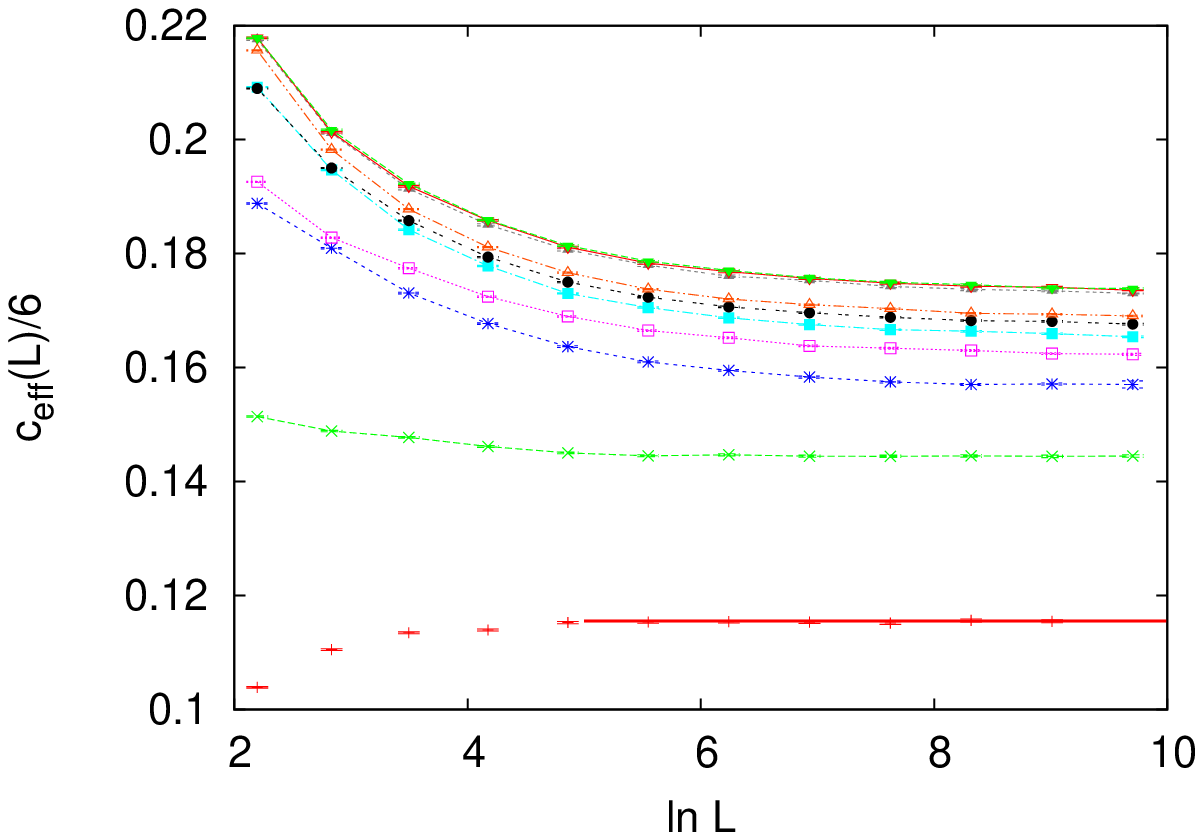} 
\caption{\label{ffull} Left. The dependence of the average entanglement entropy on the size of the subsystem in the random free-fermion model at fully connected junctions for different values of $M$. The data were obtained by the SDRG method with the maximum rule approximation.      
Right. The size-dependence of the two-point fits for the effective central charge.   
}
\end{center}
\end{figure}

Results obtained by the SDRG method for the fully connected junction can be seen in Fig. \ref{ffull}. The estimates of $c_{\rm eff}$ for various values of $M$ can be found in Table \ref{table_I} and are plotted as a function of $1/M$ in Fig. \ref{ceff}. Here, the effective central charge increases with $M$, and seems to converge to a limit, which is smaller than $3\ln 2$, for large $M$. This behavior is similar to that of the XX model, but $c_{\rm eff}$ for each $M$, as well as its limiting value are smaller than those of the XX model.  
The numerical instabilities of the exact diagonalization method for the fully connected junction appear already at smaller system sizes than for the one-dimensional model, so the method provided no reliable results.


\section{Discussion and Outlook} \label{sec:sum}

In this work, we have studied, mainly using a numerical strong-disorder renormalization group method,  the scaling of entanglement entropy in systems composed of several one-dimensional disordered chains coupled together at a multi-junction point. 

In the critical random transverse-field Ising model, the average entanglement entropy of one arm has been found to scale logarithmically with the length of the arm, and the prefactor, the effective central charge decreases with the number of arms. This tendency is in accordance with previous results on the finite-size scaling of the average magnetization $m(L)$ at the central spin \cite{stargraph}. The scaling of this quantity is given by  the probability $P(\Gamma)$ that the cluster containing the central spin has not been decimated up to the scale $\Gamma$. It has been shown that $m(L)\sim P(\Gamma_L)\sim L^{-x_M}$, where the exponent $x_M$ decreases rapidly with $M$. As entanglement between the arms is produced by decimations of the central cluster, $P(\Gamma)$ is nothing but the probability that the entanglement produced during the SDRG up to scale $\Gamma$ is zero: $P(\Gamma)={\rm Prob}[S(\Gamma)=0]$. Thus, we immediately obtain for the finite-size scaling of the probability of zero entropy: ${\rm Prob}[S(L)=0]\sim L^{-x_M}$. 
This means that, for larger $M$, the probability distribution at zero decreases slower, suggesting that the mean value increases slower with $L$. This agrees with the numerical results of the present work. 

For the XX model, we found an opposite tendency for the variation of the effective central charge: $c_{\rm eff}$ is higher for multiple junctions than in a linear chain, and increases with $M$. This can be understood since the invariant junction geometry is the fully connected one, where the actual end site of the subsystem is connected to more than one ($M-1$) sites of the environment while it has only one internal link, therefore singlet formations with external sites are more frequent for larger values of $M$.   

For the free-fermion model, the scaling of entanglement entropy depends on the connectedness of the arms, which is related to the sublattice symmetry of the model. Here, even the tendency of the variation of $c_{\rm eff}$ with $M$ can be different for different junction geometries: for star junctions $c_{\rm eff}$ decreases, while for fully connected junctions it increases with $M$. 
In the latter case, the corresponding values of $c_{\rm eff}$  are smaller than those of the XX model, in accordance with that in the decimation steps of the XX model more couplings are generated than in the free-fermion model. 
In both cases in which $c_{\rm eff}$ increases with $M$, the limiting values in the limit $M\to\infty$ are smaller than those allowed by the entanglement entropy capacity of finite chains. 

The numerical results we have presented in this work, have been obtained using a uniform distribution of the parameters of the models. The universality of the coupling distributions in the fixed point of the SDRG transformation and, as a consequence, those of the large-scale properties in linear chains is well known \cite{fisher, fisherxx}. For junction geometries, the same universality of $c_{\rm eff}$ with respect to the initial coupling distribution as for linear chains holds. The reason for this is that in the renormalization steps performed at the junction, the actual interface couplings of the arms are used, which themselves follow asymptotically the universal fixed-point distribution. This has been confirmed by numerical investigations with other types of coupling distributions.  

There are a number of ways our result could be extended and generalized. First, it would be desirable to have analytic SDRG results for the entanglement entropy asymptotics in the considered models beside the presented numerics. In particular, for the XX case we found numerically that  there is a very simple relationship between the number of arms and the central charge, Eq.~(\ref{XX_cent_charge}), and one is tempted to conjecture that at least this formula could be derived by an analytic treatment. Second, it would be instructive to investigate also other disordered models for the considered multi-arm junction geometries. It would be especially interesting to understand the effect of this geometry by examining models that are similar to each other in the one-dimensional case, but may show differences when the one-dimensional chains are glued together. One choice could be to study disordered non-interacting Kitaev wires \cite{sasa12, n17} and random-exchange Heisenberg models \cite{rm09, sykls16}, which in the one-dimensional case show a similar entanglement scaling as the random RTIM and the XX model, respectively. Third, in the considered multi-arm scenario, a natural generalizations of our results would be to calculate the entanglement between two arms. Since for non-complementary regions the entropy does not quantify entanglement anymore, one would have to choose another quantity which is an appropriate entanglement measure also for this case. The simplest choice would be the entanglement negativity,  since for  this measure both SDRG  \cite{ruggiero} and free fermion methods \cite{ez15, shapourian17, eez18} have been utilized recently.  It is clear that  for the XX and free fermion models the SDRG (random singlet state) approximation would provide an average logarithmic negativity $\mathcal{N}$ between any two arms that is related to the average entanglement entropy as $\mathcal{N}=S/(M-1)$. However, this would not hold for the disordered Ising models and Kitaev wires. Finally, our results concerning the free fermion SDRG steps also open the way for investigating this model in more general lattices. It is interesting to note that while for the one-dimensional case the disordered XX and free fermion models are equivalent, our results show that even the simplest modification of this geometry makes these models behave very differently. It would be very interesting to investigate these differences further in other lattices. In particular, already a Fano type of impurity, defined by a T shaped geometry as in Ref.~\cite{eg}, may have a very different effect in the XX and free fermion models. We leave this as a matter for future research.

\ack
We would like to thank Stefan Kehrein and Michael Knap for discussions in the early stages of this project. The work was supported by the Hungarian National Research, Development and Innovation Office (NKFIH  Grants No. K109577, K124351, K124152, and K124176) and the J\'anos Bolyai Scholarship of the Hungarian Academy of Sciences.


\end{document}